\documentstyle[12pt]{article}

\newlength{\dinwidth}
\newlength{\dinmargin}
\begin{document}

\baselineskip 10pt

\begin{center}
\begin{Large}
\begin{bf}
%
%
Fermi motion parameter $p_{_F}$ of $B$ meson
from relativistic quark model
\footnote{Talk given by D.S. Hwang, at the Int. Workshop on B-physics,
Nagoya, Japan, on Oct. 26-28, 1994. Proceedings published by World
Scientific, Singapore, edited by A. Sanda and S. Suzuki}
\\
\end{bf}
\end{Large}
\vspace{5mm}
\begin{large}
%

%
Dae Sung Hwang$^a$, C.S. Kim$^b$ and Wuk Namgung$^c$\\
\end{large}
%
%
$a$: Department of Physics, Sejong University, Seoul 133--747, Korea\\
$b$: Department of Physics, Yonsei University, Seoul 120--749, Korea\\
$c$: Department of Physics, Dongguk University, Seoul 100--715, Korea\\
\vspace{5mm}
\end{center}
\noindent
%
%
\begin{abstract}
The Fermi  motion parameter $p_{_F}$ is the most important parameter of
ACCMM model, and the value $p_{_F} \sim 0.3$ GeV has been used without clear
theoretical or experimental evidence. So, we attempted to calculate the
value for $p_{_F}$ theoretically in the relativistic quark model using
quantum
mechanical variational method. We obtained $p_{_F} \sim 0.5$ GeV, which is
somewhat larger than 0.3 GeV, and we also derived the eigenvalue of $E_B \simeq
5.5$ GeV, which is in reasonable agreement with $m_B=5.28$ GeV.
We also recalculated $|V_{ub}/V_{cb}|$ as a function of $p_{_F}$.
\end{abstract}

The simplest model for the semileptonic $B$-decay is the spectator model which
considers the decaying $b$-quark in the $B$-meson as a free particle.
The spectator model
is usually used with the inclusion of perturbative QCD radiative corrections.
The decay
width of the process $B\rightarrow X_ql\nu$ is given by
\begin{equation}
{\Gamma}_B (B\rightarrow X_ql\nu )\simeq {\Gamma}_b (b\rightarrow ql\nu )=
\vert V_{bq}{\vert}^2({{G_F^2m_b^5}\over {192{\pi}^3}})f({{m_q}\over {m_b}})
[1-{{2}\over {3}}{{{\alpha}_s}\over {\pi}}g({{m_q}\over {m_b}})],
\label{f1}
\end{equation}
where $m_q$ is the mass of the $q$-quark decayed from $b$-quark.
The decay width of the spectator model depends on $m_b^5$, therefore small
difference of $m_b$ would change the decay width significantly.

Altarelli $et$ $al.$ \cite{alta} proposed their ACCMM model for the inclusive
$B$-meson
semileptonic decays. This model incorporates the bound state
effect by treating the $b$-quark as a vitual state particle, thus giving
momentum dependence to the $b$-quark mass. The virtual state $b$-quark mass
$W$ is given by
\begin{equation}
W^2({\bf p})=m_B^2+m_{sp}^2-2m_B{\sqrt{{\bf p}^2+m_{sp}^2}}
\label{f2}
\end{equation}
in the $B$-meson rest frame, where $m_{sp}$ is the spectator quark mass,
$m_B$ the $B$-meson mass, and {\bf p} is the momentum of the $b$-quark.

For the momentum distribution of the virtual $b$-quark, Altarelli $et$ $al.$
considered the Fermi motion inside the $B$-meson with the Gaussian momentum
distribution
\begin{equation}
\phi ({\bf p})={{4}\over {{\sqrt{\pi}}p_{_F}^3}}e^{-{\bf p}^2/p_{_F}^2},
\label{f3}
\end{equation}
where the Fermi motion parameter $p_{_F}$ is treated as a free parameter.
And the decay width is given by integrating the width ${\Gamma}_b$ in
(1) with the weight $\phi ({\bf p})$. Then the lepton energy spectrum
of the $B$-meson semileptonic decay is given by
\begin{equation}
{{d{\Gamma}_B}\over {dE_l}}(p_{_F}, m_{sp}, m_q, m_B)=
{\int}_0^{p_{max}}dp\ p^2\phi ({\bf p})\
{{d{\Gamma}_b}\over{dE_l}}(m_b=W, m_q),
\label{f4}
\end{equation}
where $p_{max}$ is the maximum kinematically allowed value of $p=|{\bf p}|$.
The ACCMM model, therefore,
introduces a new parameter $p_{_F}$ for the
Gaussian momentum
distribution of the $b$-quark
inside $B$-meson instead of the $b$-quark mass of
the spectator model.
In this way the ACCMM model incorporates the bound state effects and reduces
the strong dependence on $b$-quark mass in the decay width of the
spectator model.

The Fermi motion parameter $p_{_F}$ is the most essential parameter of
the ACCMM
model as we see in the above.
However, the experimental determinations of its value
from the lepton energy spectrum have been very ambiguous until now because
various parameters, such as $p_{_F}$, $m_q$ and $m_{sp}$, are fitted
all together from the lepton energy spectrum,
and because the perturbative QCD corrections are sensitive
in the end point region of the spectrum.
We think that
the value $p_{_F} \sim 0.3$, which has been widely
used in experimental analyses, has no theoretical or experimental clean
justification, even though there has been recently an assertion
that the BSUV model~\cite{bigi} is approximately equal to ACCMM model at
$p_{_F} \simeq 0.3$.
Therefore, it is strongly required to determine the value of $p_{_F}$ more
firmly when we
think of the importance of its role in experimental analyses. The better
determination of $p_{_F}$ is also interesting theoretically since it has
its own physical correspondence related to the Fermi motion inside $B$-meson.
In this context we are
going to determine theoretically the value of $p_{_F}$ in the relativistic
quark model using quantum mechanical variational method.

We consider the Gaussian probability distribution function $\phi ({\bf p})$
in (3) as
the absolute square of the momentum space wave function $\chi ({\bf p})$ of
the bound state $B$-meson, $i.e.$,
\begin{equation}
\phi ({\bf p}) = 4\pi \vert \chi ({\bf p}){\vert}^2,
\ \ \
\chi ({\bf p}) = {{1}\over {({\sqrt{\pi}}p_{_F})^{3/2}}}
e^{-{\bf p}^2/2p_{_F}^2}.
\label{f5}
\end{equation}
The Fourier transform of $\chi ({\bf p})$ gives the coordinate space wave
function $\psi ({\bf r})$, which is also Gaussian,
\begin{equation}
\psi ({\bf r})=({{p_{_F}}\over {\sqrt{\pi}}})^{3/2}e^{-r^2p_{_F}^2/2}.
\label{f7}
\end{equation}
Then we can approach the determination of $p_{_F}$ in the framework of
quantum mechanics.
For the $B$-meson system we treat the $b$-quark non-relativistically,
but the $u$- or $d$-quark relativistically with the Hamiltonian
\begin{equation}
H=M+{{{\bf p}^2}\over {2M}}+{\sqrt{{\bf p}^2+m^2}}+V(r),
\label{f8}
\end{equation}
where $M=m_b$ is the $b$-quark mass and $m=m_{sp}$ is the $u$- or $d$- quark
mass.
We apply the variational method the Hamiltonian (\ref{f8})
with the trial wave function
\begin{equation}
\psi ({\bf r})=({{\mu}\over {\sqrt{\pi}}})^{3/2}e^{-{\mu}^2r^2/2},
\label{f9}
\end{equation}
where $\mu$ is the variational parameter.
The ground state is given by
minimizing the expectation value of $H$,
\begin{equation}
\langle H\rangle =\langle\psi\vert H\vert\psi\rangle =E(\mu ),
\ \ \
{{d}\over {d\mu }}E(\mu )=0\ \ {\rm{at}}\ \ \mu ={\bar{\mu}},
\label{f10}
\end{equation}
and then ${\bar{\mu}} = p_{_F}$ and $\bar E \equiv E({\bar{\mu}})$
approximates $m_B$.
The value of $\mu$ or $p_{_F}$ corresponds to the
measure of the radius of the two body bound state as can be seen from
$\langle r\rangle ={{2}\over{\sqrt{\pi}}}{{1}\over {\mu}}$
and
$\langle r^2{\rangle}^{{1}\over {2}} ={{3}\over {2}}{{1}\over {\mu}}$.

In (\ref{f8}) we take the Cornell potential
which is composed of the Coulomb and linear potentials,
\begin{equation}
V(r)=-{{{\alpha}_c}\over {r}}+Kr.
\label{f13}
\end{equation}
For the values of the parameters
${\alpha}_c\ (\equiv {{4}\over {3}}{\alpha}_s)$, $K$,
and the $b$-quark mass $m_b$, we use the values given
by Hagiwara $et$ $al.$~\cite{hagi},
\begin{equation}
{\alpha}_c=0.47\ ({\alpha}_s=0.35),\ \ K=0.19\ GeV^2,\ \ m_b=4.75\ GeV,
\label{f14}
\end{equation}
which have been determined by the best fit of the $(c{\bar{c}})$ and
$(b{\bar{b}})$ bound states.
For comparison we will also consider
${\alpha}_c=0.32\ ({\alpha}_s=0.24)$,
which corresponds to $\alpha_s(Q^2 = m_B^2)$.

Before applying our variational method with the Gaussian trial wave function
to the $B$-meson system, let us check the method by considering the $\Upsilon
(b{\bar{b}})$ system.
The Hamiltonian of the $\Upsilon (b{\bar{b}})$ system can be approximated
by the  non-relativistic Hamiltonian
\begin{equation}
H\simeq 2m_b+{{{\bf p}^2}\over {m_b}}+V(r).
\label{f16}
\end{equation}
With the parameters in (\ref{f14})
(or with ${\alpha}_c=0.32$),
our variational method with
the Gaussian trial wave function (\ref{f9})
gives $p_{_F}={\bar{\mu}}=1.1$ GeV and
${\bar{E}}=E({\bar{\mu}})=9.49$ GeV. Here $p_{_F}=1.1$ GeV corresponds to the
radius $R(\Upsilon )=0.2$ fm, and ${\bar{E}(\Upsilon)}=9.49$ GeV is within
$0.3\ \%$  error compared with the experimental value $E_{\rm{ exp}}=
m_{\Upsilon}=9.46$ GeV. Therefore, the
variational method with the non-relativistic Hamiltonian (\ref{f16})
gives fairly
accurate results for the $\Upsilon$ ground state.

However, since the $u$- or $d$- quark in the $B$-meson is very light, the
non-relativistic description can not be applied to the $B$-meson system.
For example, when we apply the variational method with the non-relativistic
Hamiltonian to the $B$-meson, we get the results
\begin{eqnarray}
p_{_F}=0.29\ GeV,\ \ {\bar{E}} =5.92\ GeV & & {\rm{for}}\ \
{\alpha}_s=0.35,
\label{f17}\\
p_{_F}=0.29\ GeV,\ \ {\bar{E}} =5.97\ GeV & & {\rm{for}}\ \
{\alpha}_s=0.24.
\label{f18}
\end{eqnarray}
The above masses $\bar E$ are much larger compared to the experimental value
$m_B=5.28$ GeV, and moreover the expectation values of the higher
terms in the non-relativistic perturbative expansion are bigger than those of
the lower terms. Therefore, we can not apply the variational method with the
non-relativistic Hamiltonian to the $B$-meson system.
%

Let us come back to our Hamiltonian (\ref{f8}) of the $B$-meson system.
In our variational method the trial wave
function is Gaussian both in the coordinate space and in the momentum space,
so the expectation value of $H$ can be calculated in either space from
$\langle H\rangle =\langle\psi({\bf r})\vert H\vert\psi({\bf r})\rangle
=\langle\chi({\bf p})\vert H\vert\chi({\bf p})\rangle$.
Also, the Gaussian function is a smooth function and its derivative of any
order is square integrable, thus any power of the Laplacian operator
${\nabla}^2$ is a hermitian operator at least under Gaussian functions.
Therefore, analyzing the Hamiltonian (\ref{f8}) with the variational method
can be considered as reasonable even though solving the eigenvalue equation
of the differential operator (\ref{f8}) may
be confronted with the mathematical difficulties because of the square root
operator in (\ref{f8}).

With the Gaussian trial wave function (\ref{f5}) or (\ref{f9}),
the expectation value of the Hamiltonian (\ref{f8}) can
be calculated easily besides the square root operator,
\begin{eqnarray}
\langle {\bf p}\,^2\rangle &=& \langle \psi ({\bf r}\,) |
{\bf p}\,^2| \psi ({\bf r}\,) \rangle =
\langle \chi ({\bf p}\,) | {\bf p}\,^2|
\chi ({\bf p}\,) \rangle = {3 \over 2} \mu^2,
\label{f21}\\
\langle V(r) \rangle &=& \langle \psi ({\bf r}) | -{\alpha_c \over r} + Kr \
|\psi ({\bf r}) \rangle = {2 \over \sqrt\pi} (-\alpha_c\mu + {K / \mu} ).
\label{f22}
\end{eqnarray}
Now let us consider the expectation value of the square root operator  in
the momentum space
\begin{eqnarray}
\langle \sqrt{{\bf p}\,^2+ m^2} \rangle &=& \langle \chi ({\bf p}\,)
| \sqrt{{\bf p}\,^2+ m^2} | \chi ({\bf p}\,) \rangle
= \Bigl({\mu \over \sqrt\pi}\Bigr)^3 \int_0^\infty
e^{-{p^2 / \mu^2}} \sqrt{{\bf p}\,^2+ m^2}\; d^3p
\nonumber\\
&=& {4\mu \over \sqrt\pi} \int_0^\infty e^{-x^2} \sqrt{x^2 + (m/\mu)^2} \;
x^2dx.
\label{f23}
\end{eqnarray}
The integral (\ref{f23}) can be given as a series expansion by the following
procedure. First, define
\begin{eqnarray}
I(s)\, &\equiv & \int_0^\infty \sqrt{x^2 + s} \; x^2 e^{-x^2} dx
= s^2 \int_0^\infty \sqrt{t^2 + 1} \; t^2 e^{-st^2} dt,
\label{f24}\\
I_0(s) &\equiv & \int_0^\infty \sqrt{x^2 + s} \; e^{-x^2} dx
= s \int_0^\infty \sqrt{t^2 + 1} \; e^{-st^2} dt.
\label{f25}
\end{eqnarray}
Next,  from (\ref{f24}) and (\ref{f25}),
we find the following differential relations
\begin{equation}
{d \over ds} \Bigl({I_0 \over s}\Bigr) = - {1 \over s^2} I ,
\ \ \
{d I \over d s} = - {1 \over 2} I_0 + I .
\label{f26}
\end{equation}
Combining two equations in (\ref{f26}),
we get a second order differential equation for $I(s)$,
\begin{equation}
s I''(s) - (1+s) I'(s) + {1 \over 2} I(s) = 0.
\label{f28}
\end{equation}
The series solution to (\ref{f28}) is given as
\begin{eqnarray}
I(s) &=& c_1 I_1 (s) + c_2 I_2 (s),
\nonumber\\
I_1 (s) &=& s^2 F(s; {3 \over 2}, 3) = s^2 \Bigl\{ 1 + {1 \over 2} s +
{5 \over 32} s^2 + {7 \over 192} s^3 + {7 \over 1024} s^4 + \cdots\Bigr\},
\label{f29}\\
I_2 (s) &=& I_1 (s) \int {s e^s \over [I_1 (s)]^2} ds
= - {1 \over 16} s^2 \ln s \Bigl( 1 + {1 \over 2} s + {5 \over 32} s^2 +
\cdots \Bigr)
\nonumber\\
&-&{1 \over 2} \Bigl( 1 + {1 \over 2} s + {5 \over 32} s^2 +
{7 \over 192} s^3 + {7 \over 1536} s^4 + \cdots \Bigr),
\nonumber
\end{eqnarray}
where $F(s; {3\over 2}, 3)$ is the confluent hypergeometric function which
is convergent for any finite $s$, and the integral constants $c_1\simeq-0.095$,
$c_2=-1$.
See  Appendix for the derivation of these numerical values for $c_i$.

Finally, collecting (\ref{f21}), (\ref{f22}) and (\ref{f23}),
the expectation value of $H$ is written as
\begin{eqnarray}
\langle H \rangle &=& M + {1\over 2M} \Bigl({3\over 2} \mu^2
\Bigr) + {2 \over \sqrt\pi} ( -\alpha_c \mu + K/\mu )
\nonumber\\
& & + {2\mu \over \sqrt\pi} \biggl[ 1 + {1\over 2} (m/\mu)^2 +
\Bigl({5\over 32} - 2c_1 \Bigr) (m/\mu)^4 + {1\over 4} (m/\mu)^4 \ln(m/\mu)
\biggr] ,
\label{f30}
\end{eqnarray}
up to $(m/\mu)^4$.

With  the input value of $m =m_{sp} = 0.15$ GeV, we minimize
$\langle H \rangle$ of (\ref{f30}), and then we obtain
\begin{eqnarray}
p_{_F}=\bar \mu &=& 0.54 \ GeV, \qquad m_B=\bar E = 5.54 \ GeV
\qquad {\rm for}\;
\alpha_s=0.35,
\label{f31}\\
\bar \mu &=& 0.49 \ GeV, \ \ \qquad\qquad\bar E = 5.63 \ GeV \ \qquad
{\rm for}\; \alpha_s=0.24.
\nonumber
\end{eqnarray}
Here let us check how much sensitive our calculation of $p_{_F}$ is
by considering the case where $m=m_{sp}=0$ for comparison.
For $m_{sp}=0$ the integral in (\ref{f23}) is done easily and we obtain the
following values of $\bar \mu =p_{_F}$ by the above variational method.
\begin{eqnarray}
\bar \mu &=& 0.53 \ GeV, \qquad\bar E = 5.52 \ GeV \qquad {\rm for}\;
\alpha_s=0.35,
\label{f32}\\
\bar \mu &=& 0.48 \ GeV, \qquad\bar E = 5.60 \ GeV \qquad {\rm for}\;
\alpha_s=0.24.
\nonumber
\end{eqnarray}
As we see in (\ref{f32}), the results are similar to those in (\ref{f31})
where
$m_{sp}=0.15$ GeV.
We could expect this insensitivity of the value of $p_{_F}$ to that of
$m_{sp}$ because the value of $m_{sp}$, which should be small
in any case, can not affect the integral in (23) significantly.

The calculated values of the $B$-meson mass, $\bar E$,  are much larger than
the
measured  value of 5.28 GeV.
The large values for the mass are   originated  partly because the Hamiltonian
(\ref{f30})  does not take care of the correct spin dependences  for $B$ and
$B^*$.
The difference between the pseudoscalar meson and the vector meson is  given
arise to by the
chromomagnetic hyperfine splitting, which is given by 
\begin{equation}
V_s = {2 \over 3Mm} \; \vec s_1 \cdot \vec s_2 \,\nabla^2
(- {\alpha_c \over r}).
\label{f33}
\end{equation}
Then the expectation values of $V_s$ are given by
\begin{equation}
\langle V_s \rangle = - {2 \over \sqrt\pi} \,
{\alpha_c \mu^3 \over Mm} \ \ {\rm for} \ \ B,
\ \ \
\langle V_s \rangle
= {2 \over 3\sqrt\pi} \, {\alpha_c \mu^3 \over Mm} \ \ {\rm for}
\ \ B^*,
\label{f34}
\end{equation}
and we treat $\langle V_s \rangle$ only as a perturbation.
Then we get for $B$ meson
\begin{eqnarray}
p_{_F} &=& 0.54 \ GeV, \qquad \bar E_B = 5.42 \ GeV \qquad
{\rm for}\; \alpha_s=0.35,
\label{f35}\\
p_{_F} &=& 0.49 \ GeV, \qquad \bar E_B = 5.56 \ GeV\qquad
{\rm for}\; \alpha_s=0.24 ,
\nonumber
\end{eqnarray}
and for $B^*$
\begin{eqnarray}
p_{_F} &=& 0.54 \ GeV, \qquad \bar E_{B^*} = 5.58 \ GeV \qquad
{\rm for}\; \alpha_s=0.35,
\label{f36}\\
p_{_F} &=& 0.49 \ GeV, \qquad \bar E_{B^*} = 5.65 \ GeV\qquad
{\rm for}\; \alpha_s=0.24.
\nonumber
\end{eqnarray}

The calculated values of the $B$-meson mass, 5.42 GeV ($\alpha_s = 0.35$) and
5.56 GeV ($\alpha_s = 0.24$) are in reasonable agreement compared to the
experimental value of $m_B=5.28$ GeV; the relative errors are 2.7\% and 5.3\%,
respectively.
However, for Fermi motion parameter $p_{_F}$, the calculated values, 0.54 GeV
($\alpha_s = 0.35$) and  0.49 GeV
($\alpha_s = 0.24$), are somewhat larger than the value 0.3 GeV,
which has been widely used
in the experimental
analyses of energy spectrum of semileptonic $B$-meson decay.
The value $p_{_F} = 0.3$ GeV corresponds to the $B$-meson radius
$R_B \sim 0.66$ fm,  which
seems too large. 
On the other hand, the value $p_{_F} = 0.5$ GeV corresponds to
$R_B \sim 0.39$ fm, which looks  in reasonable range.

If we use $p_{_F}=0.5$ GeV, instead of $p_{_F}=0.3$ GeV, in the experimental
analysis of the end point region of lepton energy spectrum, the value of
$|V_{ub}/V_{cb}|$ becomes significantly changed.
Using the experimental data of the end point region, {\it i.e.}
$2.3~{\rm GeV}<E_l<2.6~{\rm GeV}$ of the CLEO result \cite{cleo2},
we can find  the relation
\begin{equation}
\left| \frac{V_{ub}}{V_{cb}} \right|^2_{p_{_F}=0.5}
=
\left| \frac{V_{ub}}{V_{cb}} \right|^2_{p_{_F}=0.3}
\times
\frac{\widetilde{\Gamma}(0.3)}{\widetilde{\Gamma}(0.5)}
=\left| \frac{V_{ub}}{V_{cb}} \right|^2_{p_{_F}=0.3} \times 1.81 ,
\label{g1}
\end{equation}
where ${\widetilde{\Gamma}}(p_{_F}) \equiv
\int_{2.3}^{2.6} dE_l {\frac{d\Gamma }{dE_l}}(p_{_F})$ with $|V_{ub}|=1$,
and we assume the value of $|V_{cb}|$ is determined independently from the
other analyses.
We numerically calculated $\widetilde{\Gamma}(0.3)/\widetilde{\Gamma}(0.5)$
by using  (\ref{f4}) with $m_{sp}=0.15$ GeV, $m_q=0.15$ GeV
and $m_B=5.28$ GeV, to get its value as $1.81$.
Previously the CLEO analyzed  with
$p_{_F}=0.3$ GeV the end point lepton energy spectrum
to get \cite{cleo2}
\begin{eqnarray}
10^2 \times |V_{ub}/V_{cb}|^2&=&0.57\pm 0.11 ~~({\rm ACCMM}~ \cite{alta})
\nonumber\\
&=&1.02\pm 0.20 ~~({\rm ISGW}~ \cite{isgw}).
\label{g3}
\end{eqnarray}
As can be seen, those values are in large disagreement.
However, if we use $p_{_F}=0.5$ GeV, the result of the ACCMM model
becomes $1.03$ from (\ref{g1}), and  these two models are in good
agreement for the value of $|V_{ub}/V_{cb}|$.
Finally we show the values of $|V_{ub}(p_{_F})/V_{ub}(p_{_F}=0.3)|$ as
a function of $p_{_F}$ in Fig. 1.

\bigskip

\noindent
The work  was supported
in part by the Korean Science and Engineering  Foundation,
in part by Non-Direct-Research-Fund, Korea Research Foundation 1993,
in part by the Center for Theoretical Physics, Seoul National University,
in part by Yonsei University Faculty Research Grant,
in part by Dae Yang Academic Foundation,  and
in part by the Basic Science Research Institute Program,
Ministry of Education, 1994,  Project No. BSRI-94-2425.

\bigskip
\centerline{\bf Appendix}
\medskip

The integration constants $c_1$ and $c_2$ in (\ref{f29}) are given
by the following
relations,
\begin{eqnarray}
I(0) &=& -{1\over 2}c_2 = \int_0^\infty x^3e^{-x^2}dx = {1\over 2},
\label{a1}\\
I''(s\approx 0) &=& 2c_1 + c_2 (-{1\over 8}\ln s -{11\over 32})
\nonumber\\
&=& -{1\over 4} \int_0^\infty x^2(x^2+s)^{-3/2}e^{-x^2}dx \quad {\rm at}
\quad s\approx 0 .
\label{a2}
\end{eqnarray}
Then, from (\ref{a1}), we get
\begin{equation}
c_2\ \ =\ \ -1.
\label{a3}
\end{equation}
The integral in (\ref{a2}) can be expanded as
\begin{eqnarray}
J(s=a^2) &=&\int_0^\infty x^2(x^2+a^2)^{-3/2}e^{-x^2}dx
\nonumber\\
&=& \int_0^\infty x^2\big[(x+a)^2-2ax\big]^{-3/2}e^{-x^2}dx
\nonumber\\
&=& \int_0^\infty x^2(x+a)^{-3}\left[ 1-{2ax\over (x+a)^2} \right]^{-3/2}
e^{-x^2}dx
\nonumber\\
&=& \sum_{n=0}^\infty {(2n+1)!a^n \over 2^n(n!)^2}\int_0^\infty {x^{n+2}\over
(x+a)^{2n+3}}e^{-x^2}dx.
\label{a4}
\end{eqnarray}
Next the integral in (\ref{a4}) is obtained by successive
differentiations of an integral,
\begin{equation}
\int_0^\infty {x^{n+2}\over (x+a)^{2n+3}}e^{-x^2}dx = {1\over (2n+2)!}
\left({\partial\over\partial a}\right)^{2n+2}
\int_0^\infty {x^{n+2}\over x+a}e^{-x^2}dx.
\label{a5}
\end{equation}
Again the integral in (\ref{a5}) is related to another integral, for a small
value of $a$,
\begin{equation}
\int_0^\infty{x^{n+2}\over x+a}e^{-x^2}dx = \sum_{k=0}^{n+1}{(-a)^k \over
2} \left({n-k\over 2}\right)!+(-a)^{n+2}\int_0^\infty{e^{-x^2}\over x+a}dx.
\label{a6}
\end{equation}
The integral in (\ref{a6}) can be expanded
in a similar way as  the series
(\ref{f29}) was obtained by making use of differential equations.
For a small value of $a$,
\begin{equation}
\int_0^\infty {e^{-x^2}\over x+a} dx = -{1\over 2}e^{-a^2} (2\ln
a + \gamma + a^2 + {1\over 2} {a^4\over 2!} + \cdots  )
+ \sqrt \pi e^{-a^2} (a + {1\over 3}a^3 + {1\over 5}{a^5\over 2!} +\cdots ),
\label{a7}
\end{equation}
where $\gamma \sim 0.5772$ is the Euler's constant.
In this way the constant $c_1$ is given by an infinite series,
\begin{equation}
c_1 = -{3\over 64} + {\gamma \over 16} - {1\over 8}\sum_{n=1}^\infty
{1\over n2^n} \approx -0.0975 .
\label{a8}
\end{equation}

%
%

\vspace*{14.5cm}
\noindent
Fig.1 $|V_{ub}(p_{_F})/V_{ub}(p_{_F}=0.3)|$ as
a function of $p_{_F}$.
\vfill\eject

\end{document}